\documentclass[12pt]{article}
\pdfoutput=1
\usepackage{graphicx}
\usepackage{epsfig}
\usepackage{epstopdf}
\DeclareGraphicsExtensions{.pdf,.eps,.png,.jpg,.mps}
\usepackage{caption}
\usepackage{float}
\usepackage{color}

\usepackage{url}
\usepackage[implicit=false]{hyperref}
\usepackage{cite}

\def\lsim{\mathrel{\rlap{\lower3pt\hbox{\hskip0pt$\sim$}}
     \raise1pt\hbox{$<$}}}         
\def\gsim{\mathrel{\rlap{\lower4pt\hbox{\hskip1pt$\sim$}}
     \raise1pt\hbox{$>$}}}         

\usepackage{amsmath}
\usepackage{amsfonts}

\begin{document}
\begin{titlepage}

\centerline{\Large \bf Renormalization of the Classical Bosonic String}
\medskip

\centerline{Zura Kakushadze$^\S$$^\dag$\footnote{\, Zura Kakushadze, Ph.D., is the President of Quantigic$^\circledR$ Solutions LLC,
and a Full Professor at Free University of Tbilisi. Email: \href{mailto:zura@quantigic.com}{zura@quantigic.com}}}
\bigskip

\centerline{\em $^\S$ Quantigic$^\circledR$ Solutions LLC}
\centerline{\em 1127 High Ridge Road \#135, Stamford, CT 06905\,\,\footnote{\, DISCLAIMER: This address is used by the corresponding author for no
purpose other than to indicate his professional affiliation as is customary in
publications. In particular, the contents of this paper
are not intended as an investment, legal, tax or any other such advice,
and in no way represent views of Quantigic$^\circledR$ Solutions LLC,
the website \url{www.quantigic.com} or any of their other affiliates.
}}
\centerline{\em $^\dag$ Free University of Tbilisi, Business School \& School of Physics}
\centerline{\em 240, David Agmashenebeli Alley, Tbilisi, 0159, Georgia}
\medskip

\centerline{(May 1, 1992; in LaTeX form: November 14, 2019)\footnote{\, This note, with some (primarily, linguistic) differences, was published in 1993 in \cite{CQG}. I started working on this project prior to enrolling in a PhD program at Cornell University in November 1991 and my manuscript was received by the journal on May 1, 1992. While at Cornell University, this work was supported in part by the National Science Foundation.
}}

\bigskip
\medskip

\begin{abstract}
{}We discuss a renormalization procedure for a classical bosonic string coupled to a massless scalar field. The classical equations of motion for this system are non-renormalizable if the number of spacetime dimensions exceeds four. Our analysis indicates that string excitation modes must be included to make the string a finite object.
\end{abstract}
\medskip

\end{titlepage}

\newpage
\section{Introduction}

{}Usually, solutions of classical string equations of motion are studied in the case of
strings propagating in some dynamically-inactive background fields.
On the other hand, one can also consider strings coupled to dynamic local fields with
their kinetic terms included in the total action. In this paper
we consider the following system of a classical bosonic string coupled to a massless
scalar field:
\begin{equation}\label{action.string}
 S = -{1\over 2} \int d^2\sigma \left[T + \lambda\phi(X(\sigma))\right]h^{1/2}h^{\alpha\beta}\partial_\alpha X^\mu\partial_\beta X_\mu -{1\over 8\pi}\int d^Dx~
 \partial^\mu\phi(x)\partial_\mu\phi(x)
\end{equation}
Here $T$ is the string tension; the string is described by its $D$-dimensional spacetime coordinates $X^\mu(\sigma)$, $\mu = 1, \dots, D$,
which are defined on the worldsheet parametrized by the worldsheet coordinates
$\sigma^\alpha = (\tau, \sigma)$; $h^{\alpha\beta}(\sigma)$ is the worldsheet metric, and $h = \det(-h^{\alpha\beta})$;
the scalar field $\phi$ lives in the $D$-dimensional spacetime; $\lambda$ measures the strength of the
coupling between the scalar field and the bosonic string.

{}A similar system was considered by Dabholkar and Harvey \cite{DH}. We will study the renormalizability
of this system using the methods developed in \cite{EM} for the classical point-particle electrodynamics.
These methods can also be applied to classical equations of motion of the string, which turn out to be non-renormalizable for $D>4$.

{}The analysis in the subsequent sections allows to see in a rather
simple way that the string being an extended object, with a natural
ultraviolet cut-off at the fundamental scale $(\alpha^\prime)^{1/2}$, alone is not sufficient for its finiteness.
It is the unique combination of its excitations that provides non-renormalization properties
of the string. This combination of the local fields appearing in the effective action,
containing both massless and massive modes, is a consequence of the conformal
invariance, which is crucial for the consistency of the theory.

{}The paper is organized as follows. In Section 2 we review the renormalization of
the classical electrodynamics based on \cite{EM}. Section 3 uses these methods to renormalize
the classical string in four dimensions. In Section 4 we argue that the
theory is not renormalizable in $D > 4$. We briefly conclude in Section 5.

\section{Renormalization of Classical Electrodynamics}

Consider a relativistic point particle in $D$ dimensions coupled to the electromagnetic
field:
\begin{eqnarray}\label{S.EM}
 {\cal S} =&& -m \int d\tau \left(U^\mu(\tau) U_\mu(\tau)\right)^{1/2} - e\int d\tau \left[{\cal A}^\mu(X(\tau)) + A^\mu(X(\tau))\right] U_\mu(\tau) \nonumber\\
 &&-{1\over 16\pi}\int d^D x~ F^{\mu\nu}(x)F_{\mu\nu}(x)
\end{eqnarray}
Here $m$ is the mass of the particle; $X^\mu(\tau)$ are its coordinates on the worldline
parametrized by the proper time $\tau$; $U^\mu(\tau) = dX^\mu(\tau)/d\tau$; $e$ is the electric charge of the
particle, which is coupled to the electromagnetic field; $A^\mu(x)$ is the potential of the
field created by the particle itself; ${\cal A}^\mu(x)$ is the potential of a dynamically-inactive external electromagnetic
field. The electromagnetic field tensor for the self-field $A^\mu(x)$ is given by
\begin{equation}
 F^{\mu\nu}(x) = \partial^\mu A^\nu(x) - \partial^\nu A^\mu(x)
\end{equation}
It enters the kinetic term in the action (\ref{S.EM}). Note that the analogous quantity for the external field
\begin{equation}
 {\cal F}^{\mu\nu}(x) = \partial^\mu {\cal A}^\nu(x) - \partial^\nu {\cal A}^\mu(x)
\end{equation}
does not enter the action (\ref{S.EM}) as the external field is dynamically inactive.

{}Due to the reparametrization invariance of the action (\ref{S.EM}), we have a constraint, which can be chosen as (here $U^2(\tau) = U^\mu(\tau) U_\mu(\tau)$):
\begin{equation}
 U^2(\tau) = 1
\end{equation}
The equations of motion then read:
\begin{eqnarray}\label{self.EM}
 &&m~U^{\prime\mu}(\tau) = e\left[F^{\mu\nu}(X(\tau)) + {\cal F}^{\mu\nu}(X(\tau))\right]U_\nu(\tau)\\
 &&\partial_\nu F^{\mu\nu}(x) = -4\pi e \int_{-\infty}^{+\infty} d\tau~\delta\left(x - X(\tau)\right) U^\mu(\tau)\label{F}
\end{eqnarray}
Eq. (\ref{self.EM}) describes the self-interaction of the particle and its interaction with the external field.

{}In the Lorentz gauge $\partial^\mu A_\mu(x) = 0$, Eq. (\ref{F}) has the following solution
\begin{equation}
 A^\mu(x) = 4\pi e \int_{-\infty}^{+\infty} d\tau~G^{-}\left(x - X(\tau)\right)U^\mu(\tau)
\end{equation}
Here $G^-(x-y)$ is the retarded Green's function satisfying the following equation and boundary condition:
\begin{eqnarray}
 &&\partial^2 G^-(x-y) = \delta(x-y)\\
 &&G^-(x-y) = 0,~~~x^0 < y^0
\end{eqnarray}
We will also need another Green's function defined as:
\begin{equation}\label{combo}
 {\overline G}(x - y) = {1\over 2}\left[G^-(x-y) + G^-(y-x)\right] = {\cal G}_D\left((x-y)^2\right)
\end{equation}
where we use the fact that this Green's function depends only on the quantity $(x-y)^2$ (and
we also explicitly indicate the $D$ dependence).

{}Using (\ref{combo}), we have
\begin{eqnarray}\label{FU}
 &&F^{\mu\nu}(X(\tau))U_\nu(\tau) = 16\pi e\int_{-\infty}^\tau d\tau^\prime~{\cal G}^\prime_D\left((X(\tau)-X(\tau^\prime))^2\right) K^\mu(\tau, \tau^\prime)\\
 &&K^\mu(\tau, \tau^\prime) = \left[\left(X^\mu(\tau) - X^\mu(\tau^\prime)\right)U^\nu(\tau^\prime) - (\mu\leftrightarrow\nu)\right]U_\nu(\tau)
\end{eqnarray}

{}We can now see that the quantity $F^{\mu\nu}(X(\tau))$, i.e., the electromagnetic field tensor on the worldline, diverges.
This is due to the divergent nature of the Green's
function. Therefore, it should be regularized and the divergences, if possible, should be
removed via renormalization. From this viewpoint, renormalization
of the classical electrodynamics is analogous to renormalization in quantum
field theory: it is needed because the self-interaction of the particle is divergent. But
the analogy stops here. Thus, the renormalized equations
of motion of the classical point particle cannot be derived from the action principle.

{}The explicit form of the function ${\cal G}_D(z)$ reads:
\begin{eqnarray}
 &&{\cal G}_D(z) = \theta^{(k)}(z) / 4\pi^k,~~~D = 2k+2,~~~k=0,1,2,\dots\\
 &&{\cal G}_D(z) = (-1)^k Q^{(k)}(z) / \pi^k,~~~D = 2k+3,~~~k=0,1,2,\dots
\end{eqnarray}
where the superscript $(k)$ means the $k$-th derivative w.r.t. $z$, and
\begin{eqnarray}
 &&\theta(z) = \int_{-\infty}^z d\alpha~\delta(\alpha)\\
 &&Q(z) = (4\pi z^{1/2})^{-1}
\end{eqnarray}

{}It is convenient to use different regularizations when $D$ is even and when $D$ is
odd. When $D$ is even, we can replace the quantity $(X(\tau)-X(\tau^\prime))^2$ in (\ref{FU}) by $(X(\tau)-X(\tau^\prime))^2 - \varepsilon^2$, where $\varepsilon$ is a positive infinitesimal parameter. Then $F^{\mu\nu}(X(\tau))U_\nu(\tau)$ is equal to:
\begin{eqnarray}
 &&{\cal O}(\varepsilon),~~~D = 2\\
 &&-eU^{\prime\mu}(\tau)/2\varepsilon + {2\over 3}e\left[U^{\prime\prime\mu}(\tau) + U^\mu(\tau)~U^{\prime 2}(\tau)\right] + {\cal O}(\varepsilon),~~~D=4\label{4D}\\
 &&-eU^{\prime\mu}(\tau)/4\pi\varepsilon^3 + 3e\left[U^{\prime\prime\mu}(\tau) + 3U^\mu(\tau)~U^{\prime 2}(\tau)/2\right]^\prime/16\pi\varepsilon + \nonumber\\
 &&~~~~~~~+\mbox{finite terms},~~~D=6
\end{eqnarray}
and so forth. When $D$ is odd, we can replace the integration limit $\tau$ in (\ref{FU}) by
$\tau-\varepsilon$ and obtain for the same quantity:
\begin{eqnarray}
 &&-eU^{\prime\mu}(\tau)\ln(\rho/\epsilon) + \mbox{finite terms},~~~D=3\\
 &&-3eU^{\prime\mu}(\tau)/4\pi\varepsilon^2 + e\left[U^{\prime\prime\mu}(\tau) + U^\mu(\tau)~U^{\prime 2}(\tau)\right]/\pi\varepsilon - 3e\ln(\rho/\epsilon)\times \nonumber\\
 &&~~~~~~~\times\left[U^{\prime\prime\mu}(\tau) + 3U^\mu(\tau)~U^{\prime 2}(\tau)/2\right]^\prime/8\pi + \mbox{finite terms},~~~D=5
\end{eqnarray}
and so forth. Here $\rho$ is an arbitrary positive infrared cut-off parameter which appears
because of the logarithmic divergences.

{}So, we see that in two dimensions there is no divergence and, moreover, the point
particle does not radiate any electromagnetic waves. The reason for this is that in 2D a
free electromagnetic field is a pure gauge, although the Coulomb interaction is nontrivial.
The electromagnetic field follows the point particle preserving its constant energy and
spatial shape.

{}In three and four dimensions the self-interaction term is divergent but nonetheless
in both cases the divergences that appear have the form of the kinetic term in the initial
Lorentz equations of motion (\ref{self.EM}). Therefore, we can eliminate these divergences via
mass renormalization:
\begin{eqnarray}
 &&{\widetilde m} = m + e^2\ln(\rho/\varepsilon),~~~D=3\\
 &&{\widetilde m} = m + e^2/2\varepsilon,~~~D=4
\end{eqnarray}
Since no other divergences are present in these two cases, the theory is renormalizable.
In particular, in four dimensions using (\ref{4D}) we obtain the well-known Lorentz equation
with the radiation term \cite{LL}:
\begin{equation}
 {\widetilde m}~U^{\prime\mu}(\tau) = e {\cal F}^{\mu\nu}(X(\tau)) U_\nu(\tau) + {2\over 3}e^2\left[U^{\prime\prime\mu}(\tau) + U^\mu(\tau)~U^{\prime 2}(\tau)\right]
\end{equation}
In $D = 5,6,\dots$ we can also eliminate one divergence by renormalizing
the mass of the particle, but there are other divergences which cannot be
removed via renormalization as the required terms are absent in the
original equations of motion.

\section{Renormalization of String in Four Dimensions}

{}In this section we discuss a renormalization procedure analogous to that
in the previous section, but for the case of the four-dimensional string. The equations
of motion for the string coordinates and the scalar field following from the action (\ref{action.string}) read:
\begin{eqnarray}\label{EOM.string}
 &&\left[T + \lambda\phi(X(\sigma))\right]\partial^2X^\mu(\sigma) = \nonumber\\
 &&~~~~~~~=\lambda\left\{{1\over 2}~\partial^\mu\phi(X(\sigma))(\partial X(\sigma))^2 - \partial^\nu\phi(X(\sigma))\partial_\alpha X^\mu(\sigma)\partial^\alpha X_\nu(\sigma)\right\}\\
 &&\partial^2\phi(x) = 2\pi\lambda\int d^2\sigma~\delta(x - X(\sigma))(\partial X(\sigma))^2
\end{eqnarray}
Here the gauge freedom has been used to fix the constraint as follows:
\begin{equation}
 \partial_\alpha X^\mu(\sigma)\partial_\beta X_\mu(\sigma) = {1\over 2}~\eta_{\alpha\beta}~(\partial X(\sigma))^2
\end{equation}
So, the worldsheet is flat: $h^{\alpha\beta} = \eta^{\alpha\beta}$, where $\eta^{\alpha\beta}$ is the Minkowski metric. Note that we could also include a non-dynamical external scalar field in the action (\ref{action.string}), but this is not crucial here.

{}Using the techniques discussed in the previous section, we have (note that here we are working in four dimensions):
\begin{equation}\label{phi1}
 \phi(X(\sigma)) = \lambda\int d\sigma^\prime/|\sigma - \sigma^\prime| + \mbox{finite terms}
\end{equation}
The one-dimensional integral over the spatial coordinate $\sigma^\prime$ in (\ref{phi1}) is taken along the string. Also, the quantity
\begin{eqnarray}
 &&{1\over 2}~\partial^\mu\phi(X(\sigma))(\partial X(\sigma))^2 - \partial^\nu\phi(X(\sigma))\partial_\alpha X^\mu(\sigma)\partial^\alpha X_\nu(\sigma)=\nonumber\\
 &&~~~~~~~={\lambda\over 2}~\partial^2 X^\mu(\sigma)\int d\sigma^\prime/|\sigma - \sigma^\prime| + \mbox{finite terms}
\end{eqnarray}
can be computed using the following formula:
\begin{eqnarray}
 &&\int d\sigma^\prime\int_{-\infty}^\tau d\tau^\prime~(\sigma- \sigma^\prime)^\alpha (\sigma- \sigma^\prime)^\beta ~\delta^\prime((X(\sigma) - X(\sigma^\prime))^2) = \nonumber\\
 &&~~~~~~~=-[(\partial X(\sigma))^2]^{-2}\eta^{\alpha\beta} \int d\sigma^\prime/|\sigma - \sigma^\prime| + \mbox{finite terms}
\end{eqnarray}
Thus, the equation of motion (\ref{EOM.string}) reads
\begin{equation}
 \left\{T + {\lambda^2\over 2}\int d\sigma^\prime/|\sigma - \sigma^\prime| \right\}\partial^2 X^\mu(\sigma) = \mbox{finite nonlocal terms}
\end{equation}

{}So, the situation for the four-dimensional string is analogous to that for the relativistic point particle in $D = 3$ or $D = 4$ discussed earlier. There is
only one divergence of the same form as the kinetic term in the original equations
of motion. The logarithmically divergent integral can be regularized as follows:
\begin{equation}
 \int d\sigma^\prime/|\sigma - \sigma^\prime| = \int_{\sigma - \rho}^{\sigma-\varepsilon} d\sigma^\prime/(\sigma - \sigma^\prime) +
 \int_{\sigma+\varepsilon}^{\sigma + \rho} d\sigma^\prime/(\sigma^\prime - \sigma) = 2\ln(\rho/\varepsilon)
\end{equation}
The infrared cut-off parameter $\rho$ is analogous to that introduced in the previous section. The
renormalized string tension thus becomes:
\begin{equation}
 {\widetilde T} = T + \lambda^2\ln(\rho/\varepsilon)
\end{equation}
There is no other divergence in this case and, therefore, the string equations of motion
are renormalizable in 4D.

\section{Strings in Higher Dimensions}

{}Now we turn to the renormalizability of the string
equations of motion in higher dimensions. First note that in the string case we
need not regularize the integral over the $\tau^\prime$ variable appearing in such quantities as
$\phi(X(\sigma))$ as the extended spatial dimension of the string plays the role of the regulator for this integral. But then we
have to regularize the remaining integral over the $\sigma^\prime$ variable. So, based on the
results obtained in Section 2, we can see that in five dimensions there are two
divergences of the form $1/\varepsilon$ and $\ln(\varepsilon)$. In six dimensions we have the $1/\varepsilon^2$, $1/\varepsilon$ and $\ln(\varepsilon)$
divergences. And so forth for the higher dimensions. In 3D the string is finite.

{}The preceding analysis shows that the string is not renormalizable if $D > 4$.
In the case of the point particle
we can always eliminate one of the appearing divergences via mass
renormalization. However, in the higher-than-four-dimensional string case no divergence can be
removed via renormalization.

{}Thus, consider the leading divergences in $\phi(X(\sigma))$ and the r.h.s. of (\ref{EOM.string}) but in $D > 4$
dimensions. They enter the equations of motion in the following form:
\begin{equation}\label{HigherD}
 \propto \lambda^2 \varepsilon^{4-D} [(\partial X(\sigma))^2]^{(4-D)/2} \partial^2 X^\mu(\sigma)
\end{equation}
Therefore, this divergence cannot be removed via string tension renormalization owing to the extra factor $[(\partial X(\sigma))^2]^{(4-D)/2}$, which appears due to the following rescaling property of the Green's function in $D$ dimensions:
\begin{equation}
 {\cal G}_D(\gamma z) = \gamma^{(2-D)/2}{\cal G}_D(z)
\end{equation}
Note that for the point particle the quantity analogous to $(\partial X(\sigma))^2$ is $U^2(\tau) = 1$, which is why for the point particle the leading divergence can always be removed via mass renormalization.

\section{Concluding Remarks}

{}Let us briefly summarize the discussion in the previous sections. We have seen that
classical electrodynamics is a renormalizable theory only in $D < 5$ dimensional spacetimes.
The same holds for the bosonic string coupled to the massless scalar
field. It is worthwhile to compare our results with the work \cite{DH}, in which it was argued
that in four dimensions the superstring tension is not renormalized at all. Indeed, there
is a combination of the dilaton $\phi$, metric $g_{\mu\nu}$ and antisymmetric tensor $B_{\mu\nu}$ fields,
coupled to the bosonic string via a non-linear sigma-model action, for which the
tension of the string is not renormalized at all at the lowest order of the perturbation
theory. For the point particle there too is a certain combination of the electric charge and the mass of the
particle, for which in four dimensions the mass of the particle is not renormalized at
all due to the cancelation of the contributions from the electromagnetic and gravitational self-interactions.
However, in the case of the string we have shown that in $D > 4$ there are some other
divergences unrelated to the string tension renormalization. So, to achieve finiteness of the string, we
must include massless as well as an infinite tower of massive modes. For instance, the extra factor
$[(\partial X(\sigma))^2]^{(4-D)/2}$ in (\ref{HigherD}) in $D > 5$ indicates that there are additional modes
to be considered. This fits in the ideology of \cite{Lepage}: non-renormalizability
of a physical theory indicates that there is some new underlying physics to be included.
Our analysis of the renormalizability of classical string theory shows that the extended
nature of the string alone is insufficient for finiteness. The latter requires inclusion of
all the excitation modes of the string. On the
other hand, the way the string excitations combine together in the local field theory
action functional is determined by the conformal invariance. In particular,
the special combination of the dilaton $\phi$, metric $g_{\mu\nu}$ and antisymmetric tensor $B_{\mu\nu}$ fields
discussed above is a consequence
of the conformal invariance requirement.

\end{document}